\documentclass{article}
\usepackage{spconf,amsmath,graphicx}
\usepackage{newtxtext}
\usepackage[varg]{newtxmath}
\usepackage{subfig}
\usepackage[normalem]{ulem}
\usepackage{multirow}
\usepackage{dblfloatfix}
\usepackage{xcolor,cancel}
 
\usepackage{booktabs} 

\title{A Subjective Quality Evaluation of 3D Mesh with Dynamic Level of Detail in Virtual Reality}
\twoauthors
 {Duc V. Nguyen}
	{Tohoku Institute of Technology, Japan}
 {Tran Thuy Hien, Truong Thu Huong}
	{Hanoi University of Science and Technology, Vietnam}

\begin{document}
%
\maketitle
\begin{abstract}
3D meshes are one of the main components of Virtual Reality applications. However, a huge amount of network and computational resources are required to process 3D meshes in real-time. A potential solution to this challenge is to dynamically adapt the Level of Detail (LoD) of a 3D mesh based on the object's position and the user's viewpoint. In this paper, we conduct a subjective study to investigate users' quality perception of 3D meshes with dynamic Level of Detail in a Virtual Reality environment. The subjective experiment is carried out with five 3D meshes of different characteristics, four Levels of Detail, and four distance settings. The results of the experiment show that the impact of the dynamic level of detail depends on both the position of the 3D object in the virtual world and the number of vertices of the original mesh. In addition, we present a quality model that can accurately predict the MOS score of a LoD version of a 3D mesh from the number of vertices and the distance from the viewpoint.
\end{abstract}
\section{Introduction}

 3D models play a central role in constructing a Virtual Reality environment. For instance, 3D models of the museum itself and all artifacts are needed to build a VR-based virtual museum. A 3D model can be represented in several formats, notably polygon mesh and point cloud. A polygon mesh (3D mesh)~\cite{PolygonMesh} represents a 3D model by using polygons, usually triangles, that form a network of faces. Each polygon is represented by vertices, edges, and a normal vector that defines its orientation. The surface of the model can be smoothed, textured, or colored by applying different attributes to the polygon and its vertices. A point cloud is a collection of points in space that defines the shape and surface of the 3D model~\cite{PointCloud}. Each point is attributed by its coordinates, color, and intensity. In this paper, we consider 3D models represented by 3D meshes.


A Virtual Reality environment usually contains a large number of objects. Thus, the system must be able to process a large number of 3D meshes in real time to offer users a high quality of experience. However, 3D mesh processing is a resource-intensive task, requiring a large amount of not only computational resources but also network resources. The computational resource is mainly associated with the rendering of a 3D mesh, which generates a 2D image from the 3D mesh given the user's viewpoint. In general, complex 3D meshes demand more computational resources than simpler ones. The network resource is required if the 3D meshes are stored on a remote server as in the case of Web-based Virtual Reality~\cite{quang2023visibility}. In this case, the user agent (i.e., the web browser) must download all the 3D meshes stored on the server to the user device. Like the computational resources, complex 3D meshes have large data sizes and thus demand more network resources than simple 3D meshes.



Dynamic Level of Detail (LoD) is a potential method to optimize the transmission and processing of 3D meshes on resource-constrained devices. In particular, multiple versions of different levels of detail are generated from the original mesh in advance. The system then decides an appropriate LoD version for each 3D mesh based on the object's position and the user's viewpoint~\cite{quang2023visibility}. In general, high LoD versions are chosen for those models that are close to the viewer. On the other hand,  versions with low LoD are selected for 3D meshes that are far away.  

There are two key questions in dynamic Level of Detail-based VR systems, which are:
\begin{enumerate}
    \item How to generate LoD versions of a 3D mesh?
    \item How to select the appropriate LoD version of a 3D mesh given the system constraints?
\end{enumerate}
So far, existing works mainly focus on answering the second question by proposing various LoD version selection algorithms~\cite{luebke2003level,Distance-LoD,quang2023visibility}. To answer the first question, it is critical to understand the user's perception of a LoD version in a Virtual Reality environment. A LoD version of a 3D mesh can be generated by several methods such as downsampling, geometric simplification, and compression. In this paper, we consider geometric simplification as the LoD version generation method and leave the other methods for our future work. Our investigation shows that the effects of geometry simplification are largely dependent on the number of vertices of the original meshes. In particular, for 3D meshes with 100K+ vertices, it is possible to remove up to 95\% of the vertices without significant loss in user-perceived quality. Based on the subjective experiment results, we present a non-reference quality model for predicting the MOS score of a LoD version based on the distance from the viewer and the number of vertices of the LoD version.

The rest of this paper is organized as follows. Section 2 gives an overview of the related work. Section 3 describes our experiment settings. Section 4 presents the result analysis.  Section 5 describes the quality model for predicting MOS score of a given LoD version. Finally, Section 6 concludes this paper.

\section{Related work}

In the literature, dynamic Level of Details has been applied to optimize the transmission and rendering of 3D mesh~\cite{luebke2003level,QuangMMasia2023,Petrangeli2019}. In particular, multiple versions with different Levels of Details of the original 3D mesh are generated using techniques such as downsampling~\cite{quang2023visibility}, edge collapse~\cite{hoppe1996progressive}, and compression~\cite{Hooft2019,QuangMMasia2023}. Then, the suitable LoD version for each 3D mesh is chosen to maximize the user quality of experience given the constraints on network and processing resources. For LoD version selection, previous works rely on objective quality metrics such as screen-space error~\cite{luebke2003level,Petrangeli2019}, and projected screen area~\cite{quang2023visibility,QuangMMasia2023} to decide the importance of each 3D mesh as well as the overall quality of all 3D meshes presented in the scene, then ranking-based algorithm~\cite{Hooft2019} or dynamic programming-based  algorithms~\cite{Petrangeli2019,QuangMMasia2023} are applied to find the appropriate LoD version for each 3D mesh.

The user's quality perception of a 3D mesh is affected by various factors including content-related factors, environment-related factors, and processing-related factors. Previous works have studied the impacts of content-related factors such as content characteristics~\cite{nehme2023textured}, and diffuse colors~\cite{Yana2021}. The impacts of processing-related factors such as distortion types~\cite{pan2005quality,Guo2016}, and deformation interaction~\cite{nehme2023textured} have also been investigated. Previous works have also investigated the influence of environment-related factors such as lighting condition~\cite{gutierrez2020quality}, and light-material interaction~\cite{Vanhoey2017}. The impact of the distance from the viewpoint is first studied in our previous work~\cite{DucIEICETrans2024}. However, the subjective experiment in~\cite{DucIEICETrans2024} is carried out in a non-virtual reality environment, and thus the findings in that work can not be applied to Virtual Reality. 

\begin{figure*}[t]
    \begin{minipage}[b]{0.18\linewidth}
        \centering
        \includegraphics[width=\linewidth]{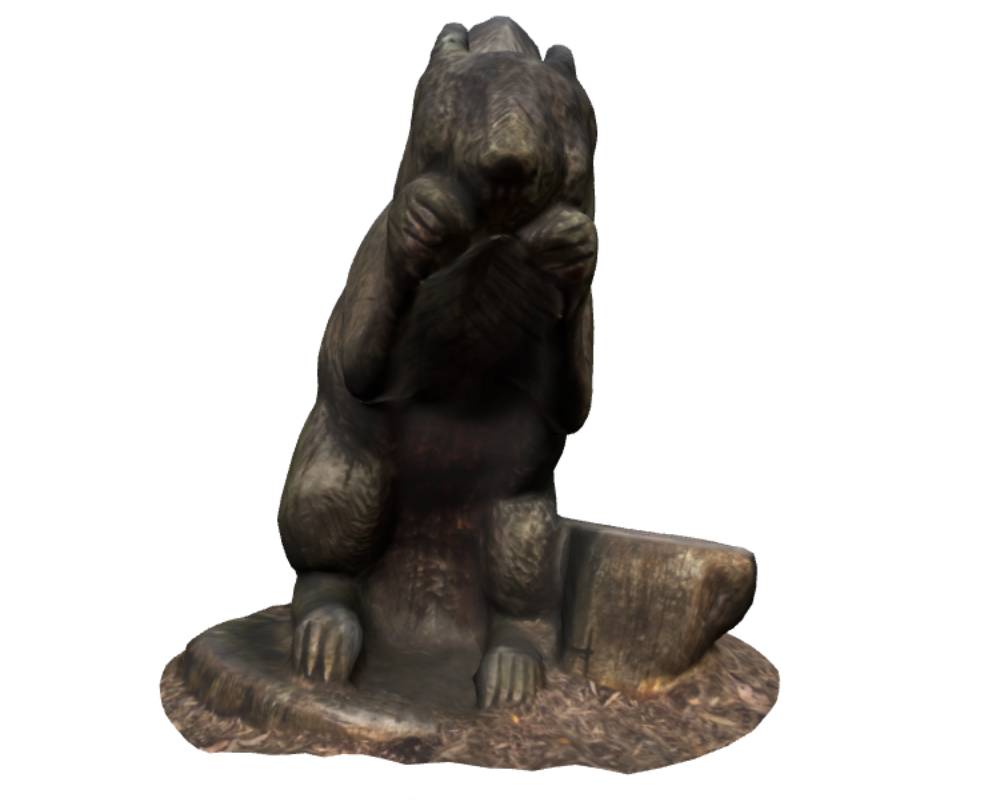}
        \caption*{Squirrel}
    \end{minipage}
    \hfill
    \begin{minipage}[b]{0.18\linewidth}
        \centering
        \includegraphics[width=\linewidth]{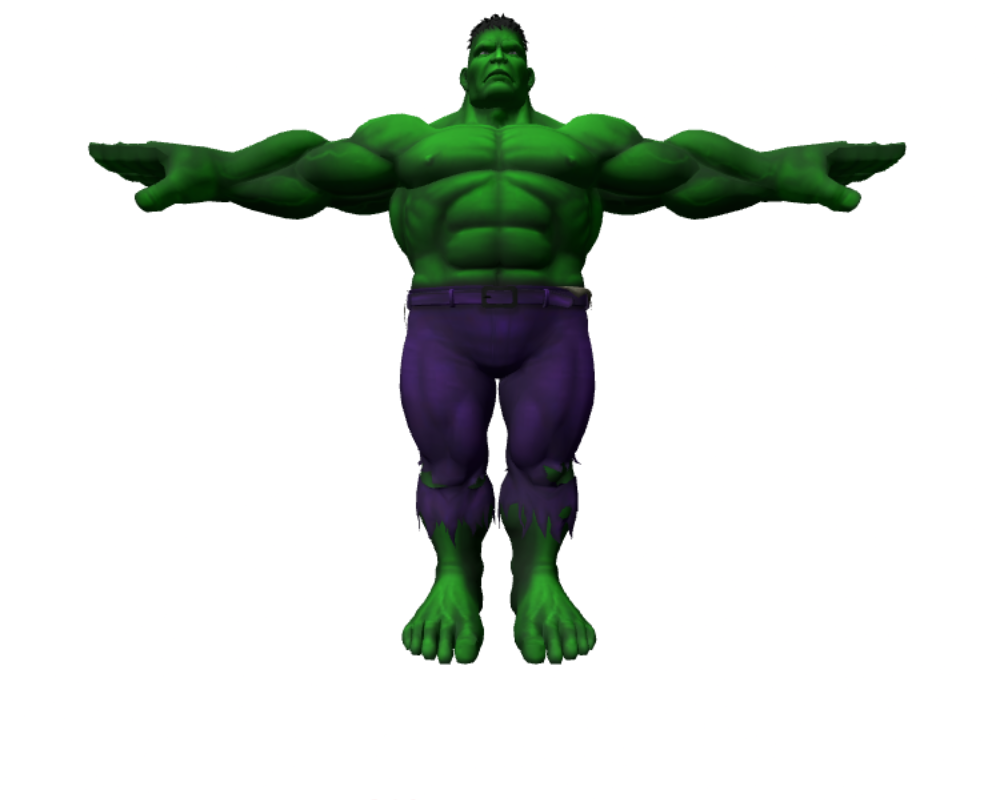}
        \caption*{Hulk}
    \end{minipage}
    \hfill
    \begin{minipage}[b]{0.18\linewidth}
        \centering
        \includegraphics[width=\linewidth]{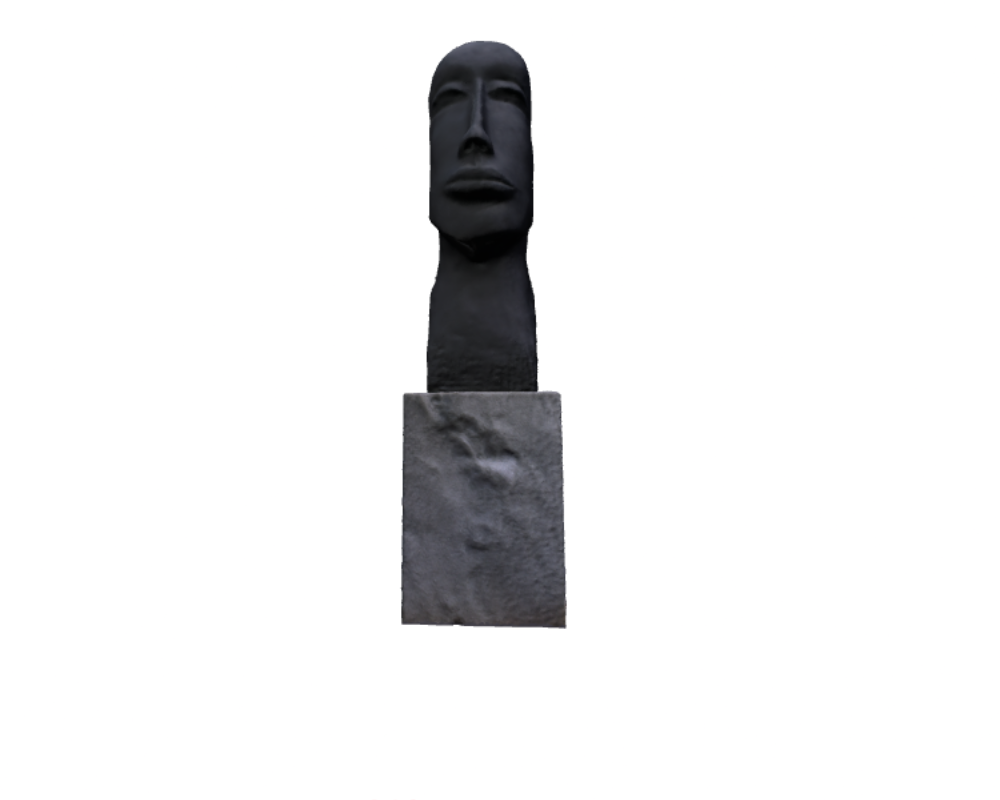}
        \caption*{Statue}
    \end{minipage}
    \hfill
    \begin{minipage}[b]{0.18\linewidth}
        \centering
        \includegraphics[width=\linewidth]{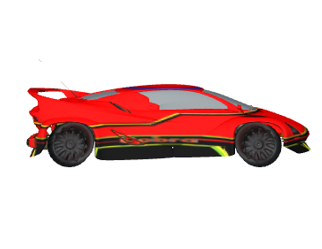}
        \caption*{Car}
    \end{minipage}
    \hfill
    \begin{minipage}[b]{0.18\linewidth}
        \centering
        \includegraphics[width=\linewidth]{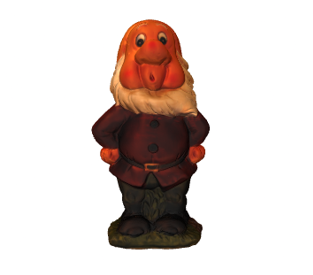}
        \caption*{Dwarf}
    \end{minipage}
    \caption{Snapshots of five 3D meshes used in our subjective experiment. The models are obtained from the public dataset constructed in~\cite{Guo2016}.}
    \label{fig:3d_mesh}
\end{figure*}
\begin{table*}[t]
\caption{Number of vertices (size) of four LoD versions of the five 3D meshes.}
\centering
\label{tab:vertex_num}
\begin{tabular}{cccccl}
\toprule
\textbf{LoD Version} & \textbf{Squirrel} & \textbf{Hulk} & \textbf{Statue} & \textbf{Car} & \textbf{Dwarf}\\ 
\midrule
Version 1 & 3099 (581KB) & 5257 (953KB) & 52002 (10822KB) & 62060 (12137KB) & 125004(26587KB)\\ 
Version 2 & 1865 (345KB) & 3241 (577KB) & 31195 (6452KB) & 37665(7290KB)& 75004(15659KB)\\ 
Version 3 & 629 (111KB) & 1187 (195KB) & 10389 (2076KB) & 13087(2441KB)& 25004(5235KB)\\ 
Version 4 & 318 (56KB) & 668 (99KB) & 5187 (1017KB) & 6890(1235KB) & 12504(2601KB) \\ 
\bottomrule
\end{tabular}
\end{table*}

\section{Subjective Quality Experiment}
\subsection{Experiment Conditions}
In this study, we use five 3D meshes from the public dataset constructed in~\cite{Guo2016}. The snapshots of five 3D meshes are shown in Fig.~\ref{fig:3d_mesh}. The Squirrel and Statue models are reconstructed from multiple images of actual objects, so the texture images of these meshes are noisier and contain more complex texture seams. The Dwarf model is a scanned model created from reconstruction and scanning. The Hulk and Car models are artificial models created using modeling software with structured texture content and smooth textured seams. These 3D meshes cover a wide variety of texture and geometry properties. The number of vertices ranges from 6K (Squirrel model) to 250K (Dwarf model).

For the subjective experiment, we apply simplification of geometric data to create LoD versions of the 3D mesh as proposed in~\cite{quang2023visibility}. In particular, the 3D meshes are simplified based on iterative edge contraction and quadratic error metrics, which can rapidly produce high-quality approximations of such models. This simplification does not access the pixels of the texture, which merely tries to update the texture coordinates of the texture map so that the texture is mapped onto the surface in the same way as the original model. Four LoD versions are generated for each 3D mesh with the number of vertices that are 50\%, 30\%, 10\%, and 5\% of the number of vertices of the original model using the MeshLab software~\cite{MeshLab}. The number of vertices and the sizes in KBytes of the LoD versions of five 3D meshes are shown in Table~\ref{tab:vertex_num}. It is worth noting that the size of the LoD versions is proportional to the number of vertices. Also, deciding the optimal number of vertices to be retained for each LoD version is an interesting research question and is reserved for our future work. 


\begin{figure}[t]
    \centering
    \subfloat[Reference mesh (left) and impaired mesh (right)]
    {
        \centering
        \includegraphics[width=0.5\columnwidth]{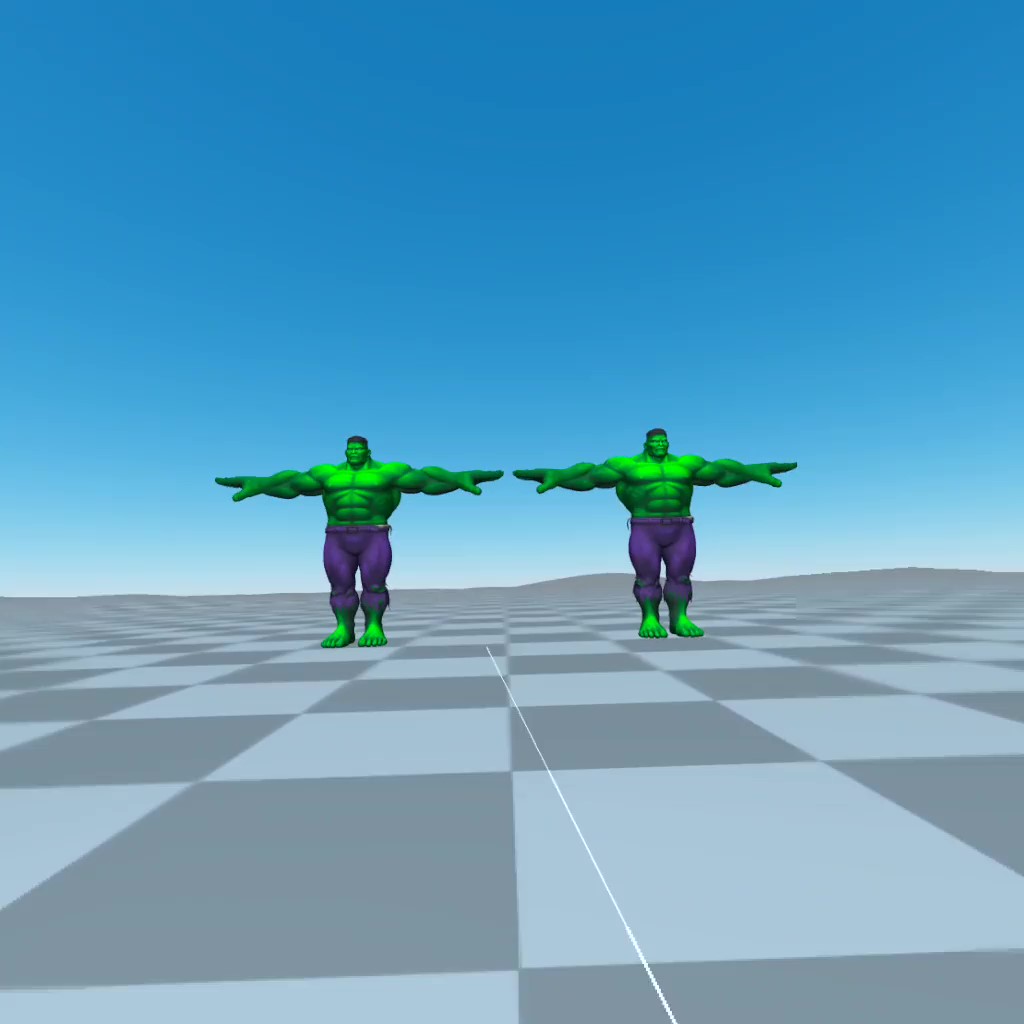}
        \label{fig:test_env_1}
    }
    \subfloat[Scoring window. Participants give scores using the controller.]
    {
        \centering
        \includegraphics[width=0.5\columnwidth]{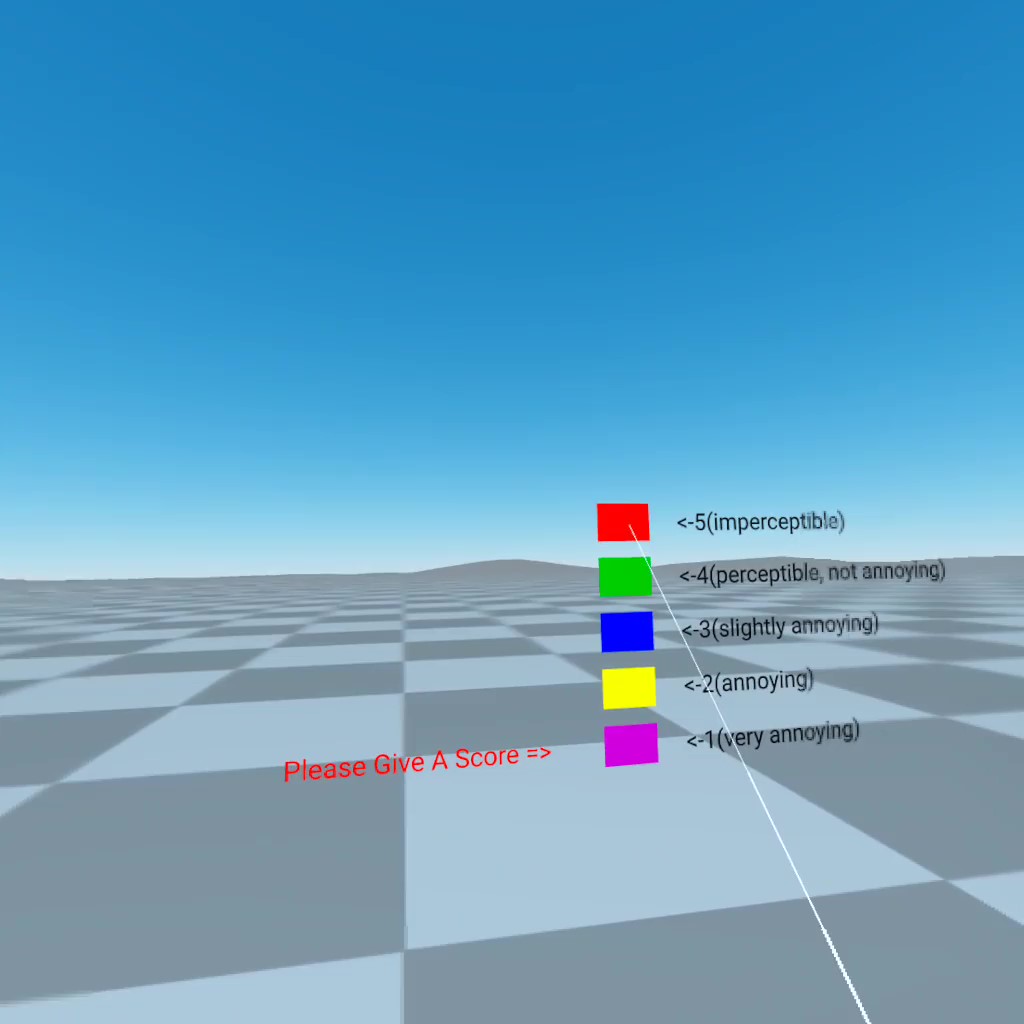}
        \label{fig:test_env_1}
    }
    \caption{Subjective test environment.}
    \label{fig:test_env}
\end{figure}

\subsection{Procedure}
To carry out the subjective experiment, we develop a test environment using A-Frame~\cite{AFrame}, which is a Web-based Virtual Reality framework. In the test environment, the position and direction of the viewpoint are fixed while the location of a LoD version of a 3D mesh is varied. In this study, we consider four distances between the viewpoint and the 3D meshes which are $d=\{5, 10, 15, 20\}$ meters. As a result, our experiment consists of 80 stimuli.

The LoD versions are in Object file format and are loaded at the beginning of each test session. The stimuli are displayed in random order and participants are asked to give their opinion scores on a 5-grade scale. We utilize the double-stimulus impairment scale (DSIS) approach to determine the subjective score for each stimulus~\cite{bt2020methodologies}. In this method, the LoD version and the original 3D mesh are displayed side by side with which the LoD version is displayed on the right side of the viewport as shown in Fig.~\ref{fig:test_env}a.

To allow the participant to accurately access the visual quality of a 3D mesh, a LoD version and the corresponding original 3D mesh are rotated clockwise three times during the test. The time interval between rotations is set to 4s, so the total time of appearance for a LoD version is approximately 12s. After that, the viewer is asked to rate the quality of the impaired mesh on a 5-grade scale as follows: 5 (imperceptible), 4 (perceptible, but not annoying), 3 (slightly annoying), 2 (annoying), and 1(very annoying). The scores are displayed in the test environment as shown in Fig.~\ref{fig:test_env}b. There is no time limit to vote, and the stimuli are not shown during the voting period. The subjective study is conducted in a Virtual Reality setting using an Oculus Quest 2 headset, with a resolution of 1832 × 1920 per eye and a refresh rate of 90 Hz. The participants use controllers to give scores for each stimulus.

A total of 20 participants took part in the subjective experiment, aged between 19 and 32, all with normal or corrected-to-normal vision. At the beginning of each test session, we explain the objectives of the test and the test procedure to the participant. On average, it takes about 30 minutes for each participant to finish the experiment. The Mean Opinion Score (MOS) of a stimulus is calculated as the average score of all participants who have evaluated the stimulus.

\begin{figure}[t]
    \centering
    \includegraphics[width=\columnwidth]{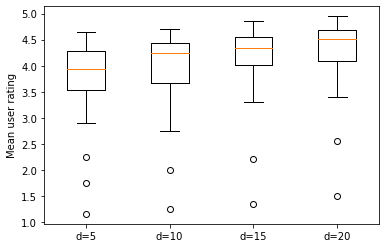}
    \caption{Mean user ratings at five considered distances}
    \label{fig:MOS_distance_boxplot}
\end{figure}

\begin{table}[t]
\caption{Results of the one-way ANOVA test.}
\label{tab:one-way-ANOVA}
\resizebox{\columnwidth}{!}{%
\begin{tabular}{|l|l|l|l|l|l|l|}
\hline
\textbf{Source   of Variation} & \textbf{SS} & \textbf{df} & \textbf{MS} & \textbf{F} & \textbf{P-value} & \textbf{F crit} \\ \hline
Between Groups & 36.43269 & 4 & 9.108172 & 23.20878 & 1.66E-12 & 2.493696 \\ \hline
Within Groups & 29.43339 & 75 & 0.392445 &  &  &  \\ \hline
\end{tabular}%
}
\end{table}

\begin{figure*}[t]
    \subfloat[Squirrel]{
        \centering
        \includegraphics[width=0.33\textwidth]{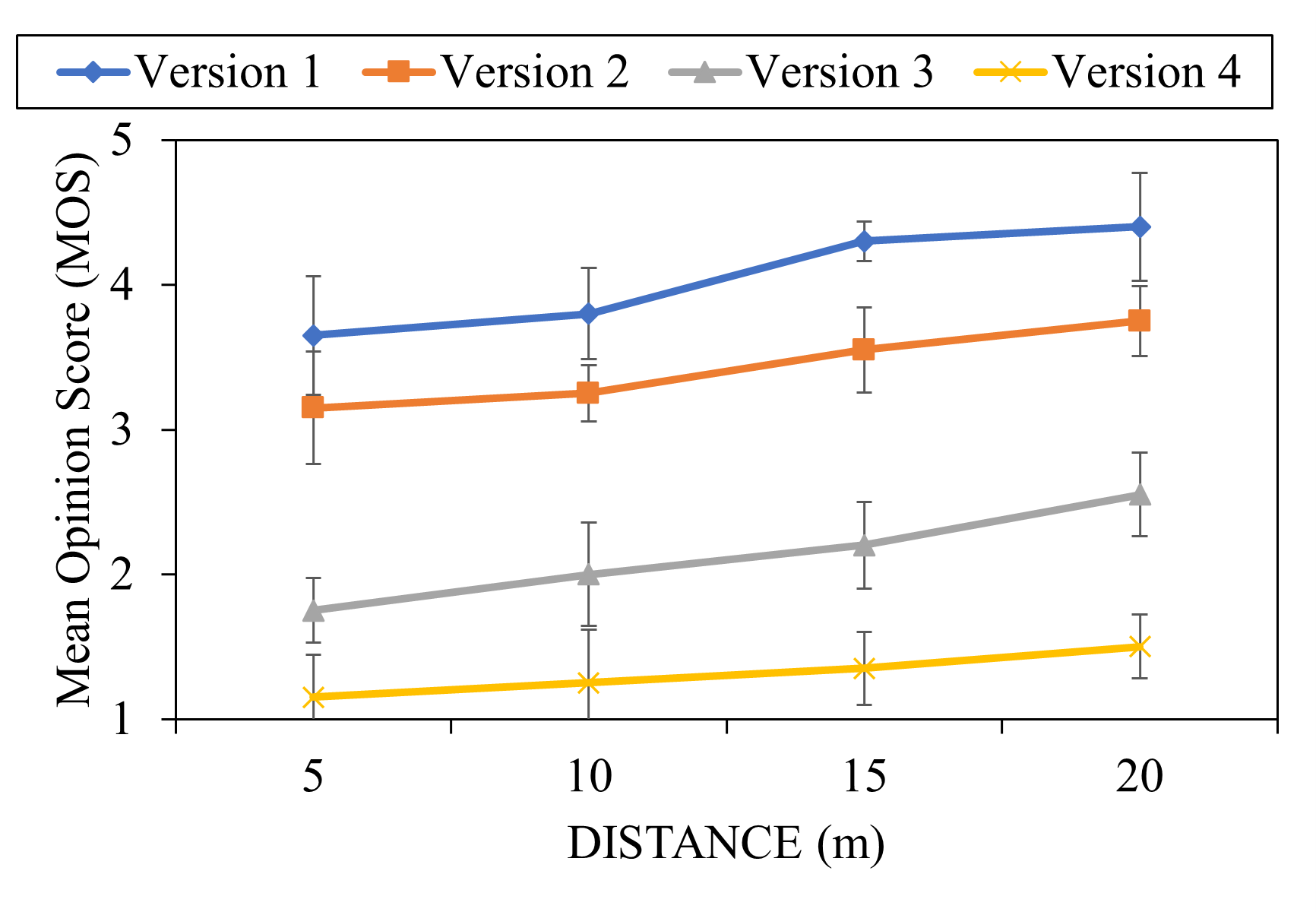}
    }
    \subfloat[Hulk]{
        \centering
        \includegraphics[width=0.33\textwidth]{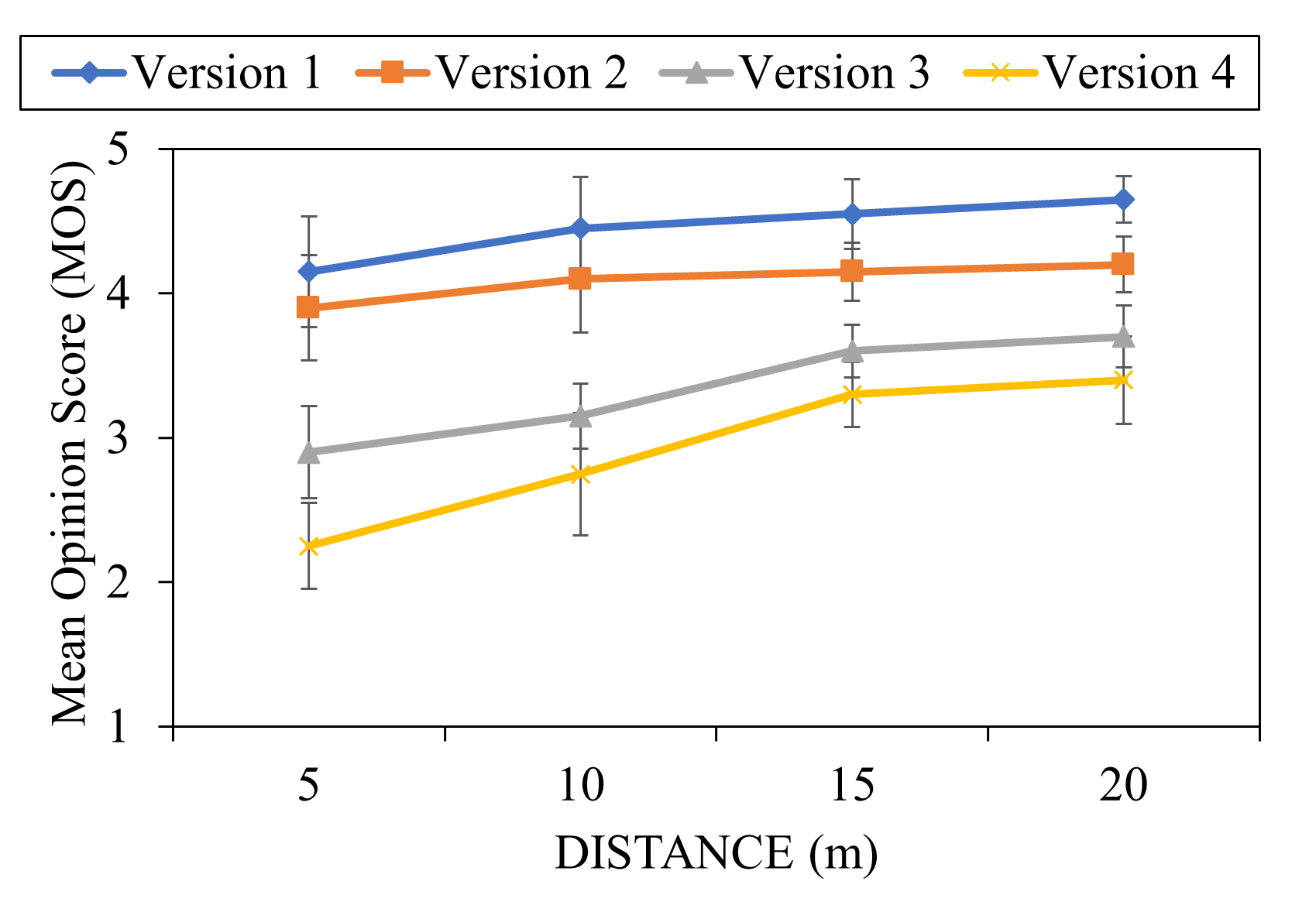}
    }
    \subfloat[Statue]{
        \centering
        \includegraphics[width=0.33\textwidth]{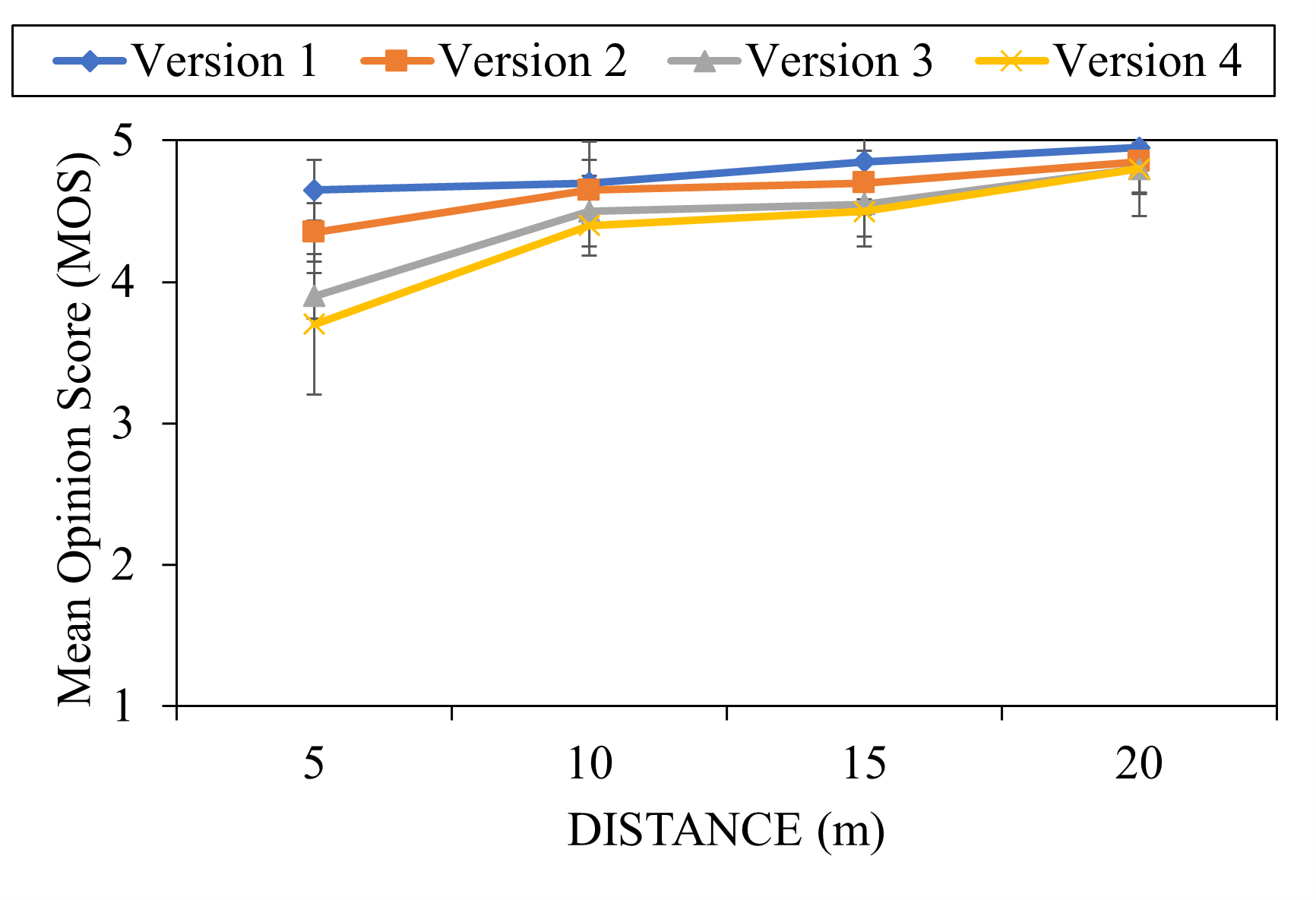}
    }
    \hfill
    \subfloat[Car]{
        \centering
        \includegraphics[width=0.33\textwidth]{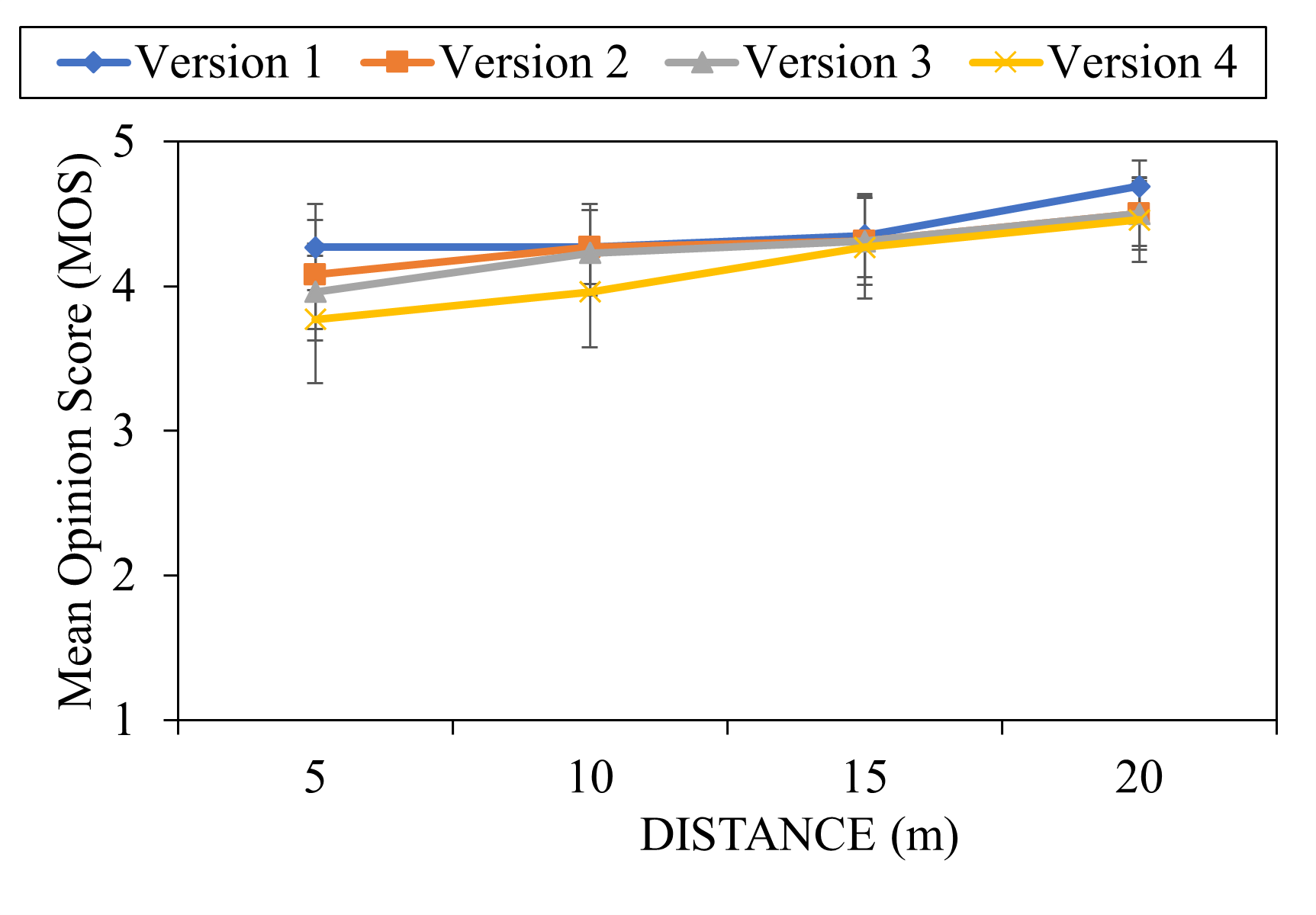}
    }
    \subfloat[Dwarf]{
        \centering
        \includegraphics[width=0.33\textwidth]{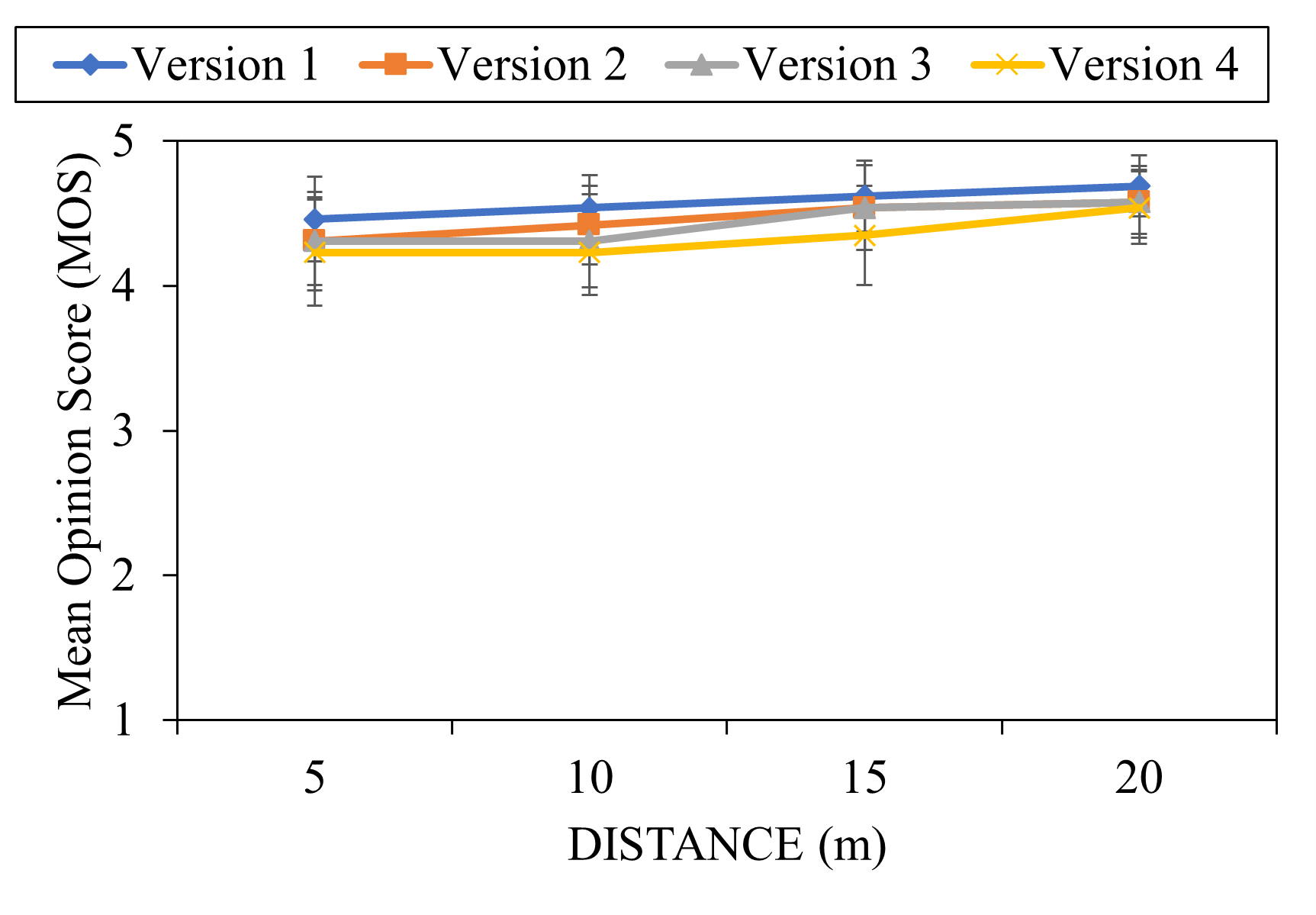}
    }
    \caption{MOSs of the LoD versions of five 3D meshes as the distance from the viewpoint increases.}
    \label{fig:MOS}
\end{figure*}

\section{Experiment Results and Analysis}
In this section, we present and analyze the results of the subjective experiment described in the previous section.

\subsection{Statistical Analysis}
First, we conduct a statistical analysis of the obtained user ratings. The results of the one-way ANOVA test are shown in Table~\ref{tab:one-way-ANOVA}. It can be seen that the p-value is less than 0.05, indicating that there is a statistically significant difference in the user ratings of the five 3D meshes. We additionally conduct a follow-up test to investigate the statistical difference between pairs of 3D meshes. The results of the Tukey's HSD Pairwise Group Comparisons test with 95.0\% Confidence Interval show that there is a statistically significant difference between \textit{Squirrel} and \textit{Hulk}, \textit{Squirrel} and \textit{Statue}, \textit{Squirrel} and \textit{Car}, Squirrel and \textit{Dwarf}, \textit{Hulk} and \textit{Statue}, and \textit{Hulk} and \textit{Dwarf} ($p < 0.05$). Meanwhile, there is no statistically significant difference in the user ratings between \textit{Hulk} and \textit{Car}, \textit{Statue} and \textit{Car}, \textit{Statue} and \textit{Dwarf}, and \textit{Car} and \textit{Dwarf} ($p > 0.05$).  

\subsection{Impact of Distance}
Next, we analyze the impact of the distance on the MOS scores of the LoD versions of five considered 3D meshes. Fig.~\ref{fig:MOS_distance_boxplot} shows the boxplot of the user ratings at five considered distances. It can be seen that the mean user rating increases as the distance between the 3D mesh and the viewpoint increases. This result indicates that the further the distance is, the more difficult for the participants to recognize the distortion level of a LoD version. This result can be explained by the fact that the size of a 3D mesh on the viewport becomes smaller as the distance increases. 

Fig.~\ref{fig:MOS} shows the MOSs of the LoD versions of the five 3D meshes at four distance values with 95\% confident interval. It can be seen that the impact of the distance varies significantly across LoD versions and 3D meshes. Table~\ref{tab:QoE-increase} shows the increases in MOS score as the distance increases from $d=5m$ to $d=20m$ of the LoD versions of the five 3D meshes. For the \textit{Squirrel} model, the MOS values of the LoD versions increase by 0.35$\sim$0.8 as the distance increases from $d=5$ to $d=20$. Among the LoD versions, Version 4 has the smallest amount of increase while Version 3 has the biggest. The MOS scores of Version 4 of \textit{Hulk} and \textit{Statue} models are increased significantly by more than 1 MOS. For the \textit{Statue}, \textit{Car}, and \textit{Dwarf} models, LoD versions with a smaller number of vertices have a bigger amount of increase in MOS score than those with less number of vertices. It can be also seen that the differences in the amount of increase in MOS between the LoD versions are relatively small for the \textit{Car} and \textit{Dwarf} models. Especially, the difference between Version 1 and Version 4 is less than 0.1 MOS.

\begin{table}[t]
\caption{Amount of increase in MOS of the LOD versions of five 3D meshes as the distance increases from $d=5$ to $d=20$. Bold numbers indicate the biggest while underscored bold numbers increase the smallest.}
\label{tab:QoE-increase}
\resizebox{\columnwidth}{!}{%
\begin{tabular}{|l|c|c|c|c|c|}
\hline
LoD Version & \multicolumn{1}{l|}{\textbf{Squirel}} & \multicolumn{1}{l|}{\textbf{Hulk}} & \multicolumn{1}{l|}{\textbf{Statue}} & \multicolumn{1}{l|}{\textbf{Car}} & \multicolumn{1}{l|}{\textbf{Dwarf}} \\ \hline
Version 1 & 0.75 & 0.5 & { \textbf{0.3}} & { \textbf{0.42}} & { \textbf{0.23}} \\ \hline
Version 2 & 0.6 & { \textbf{0.3}} & 0.5 & { \textbf{0.42}} & 0.27 \\ \hline
Version 3 & \textbf{0.8} & 0.8 & 0.9 & 0.54 & 0.27 \\ \hline
Version 4 & { \textbf{0.35}} & \textbf{1.15} & \textbf{1.1} & \textbf{0.69} & \textbf{0.31} \\ \hline
\end{tabular}%
}
\end{table}
\begin{figure*}[t]
    \subfloat[Squirrel]{
        \centering
        \includegraphics[width=0.33\textwidth]{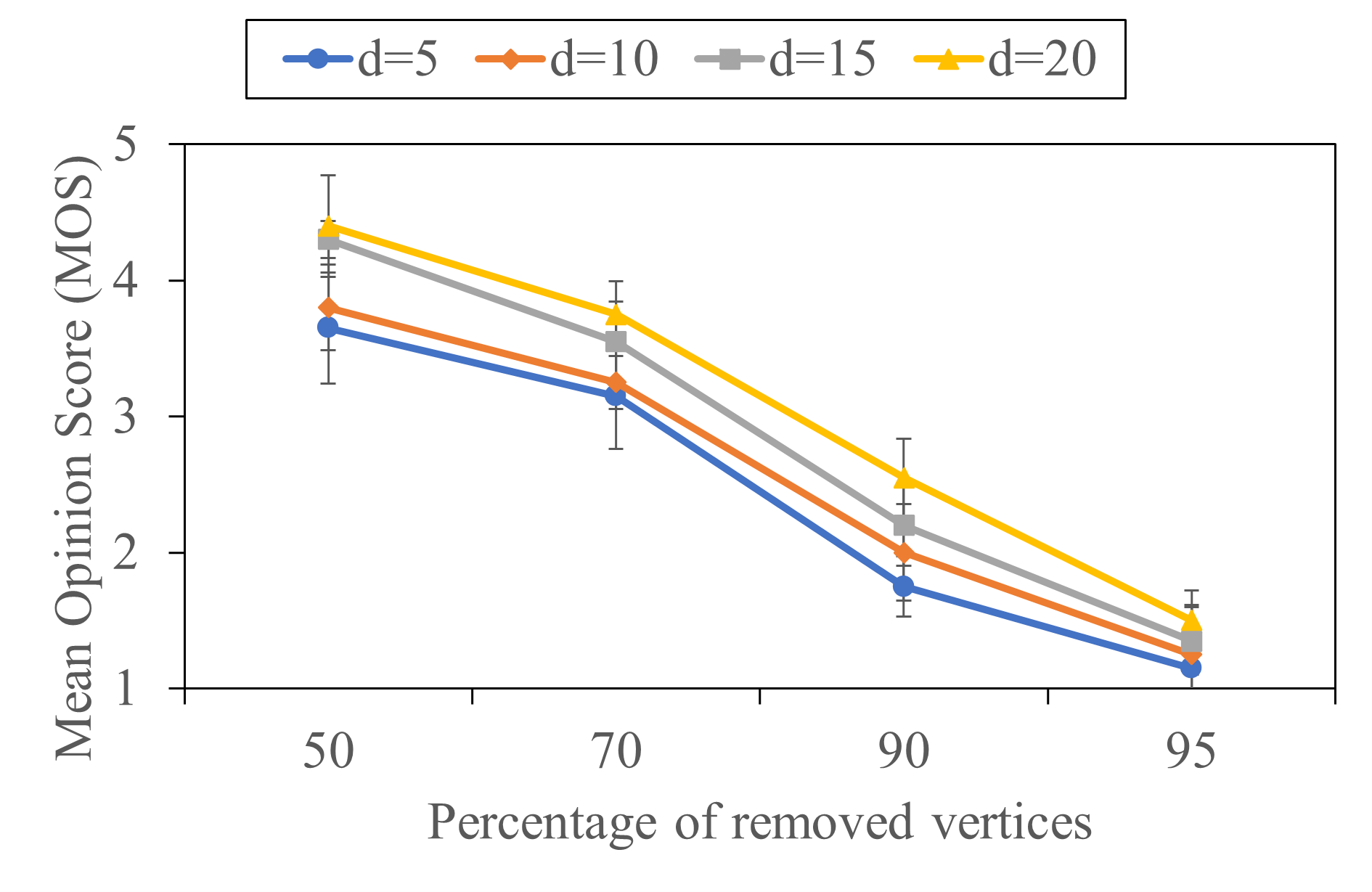}
    }
    \subfloat[Hulk]{
        \centering
        \includegraphics[width=0.33\textwidth]{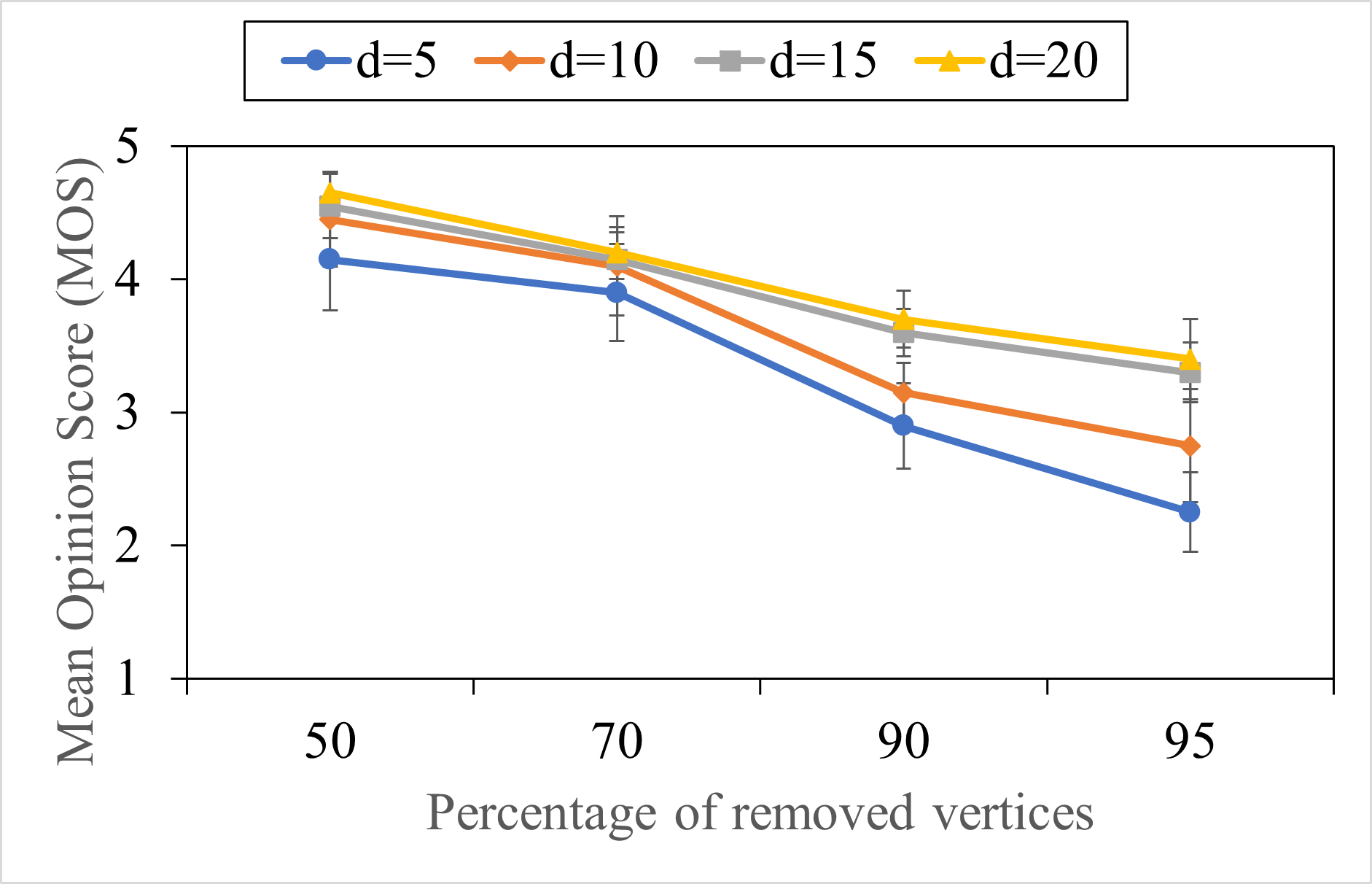}
    }
    \subfloat[Statue]{
        \centering
        \includegraphics[width=0.33\textwidth]{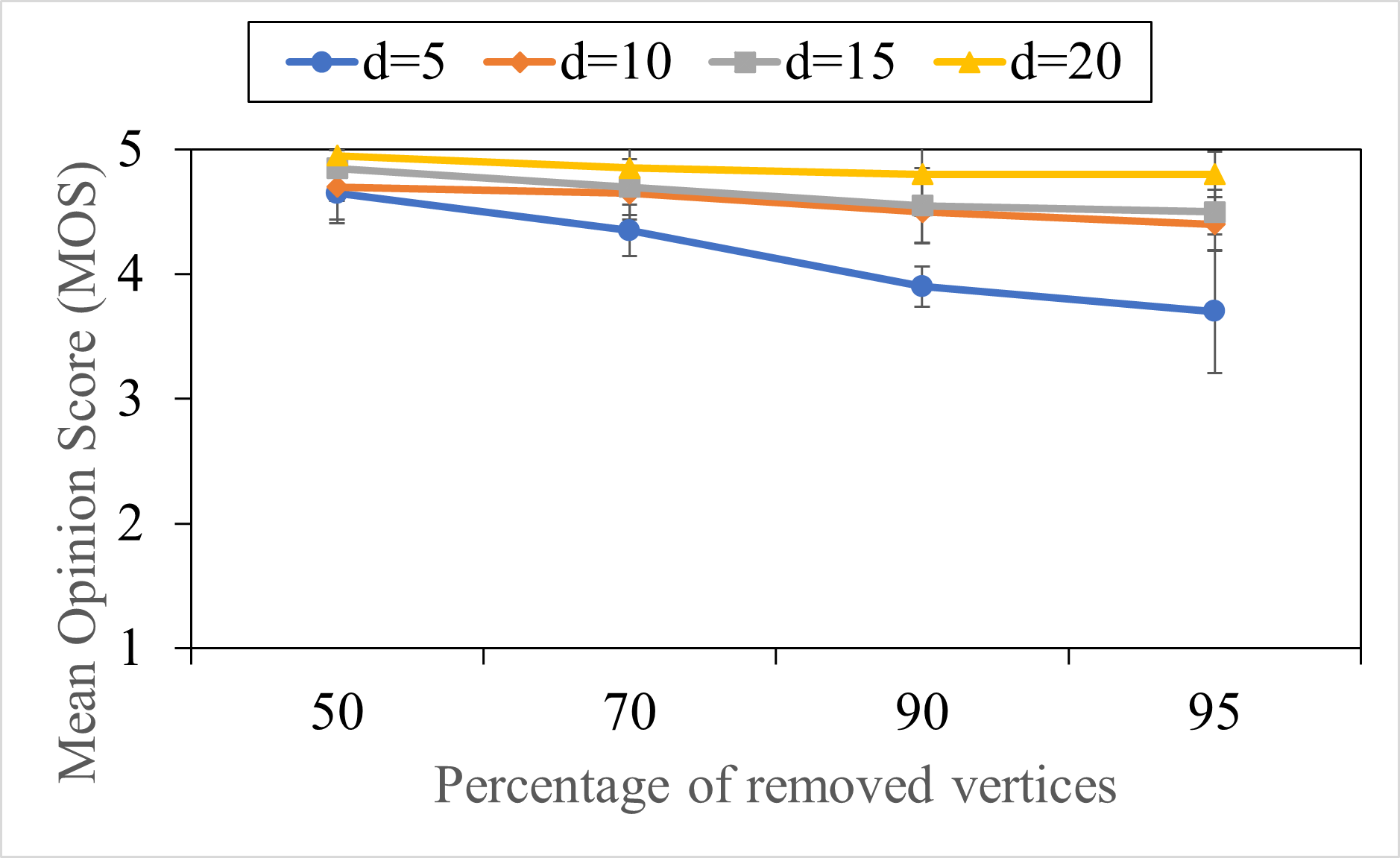}
    }
    \hfill
    \subfloat[Car]{
        \centering
        \includegraphics[width=0.33\textwidth]{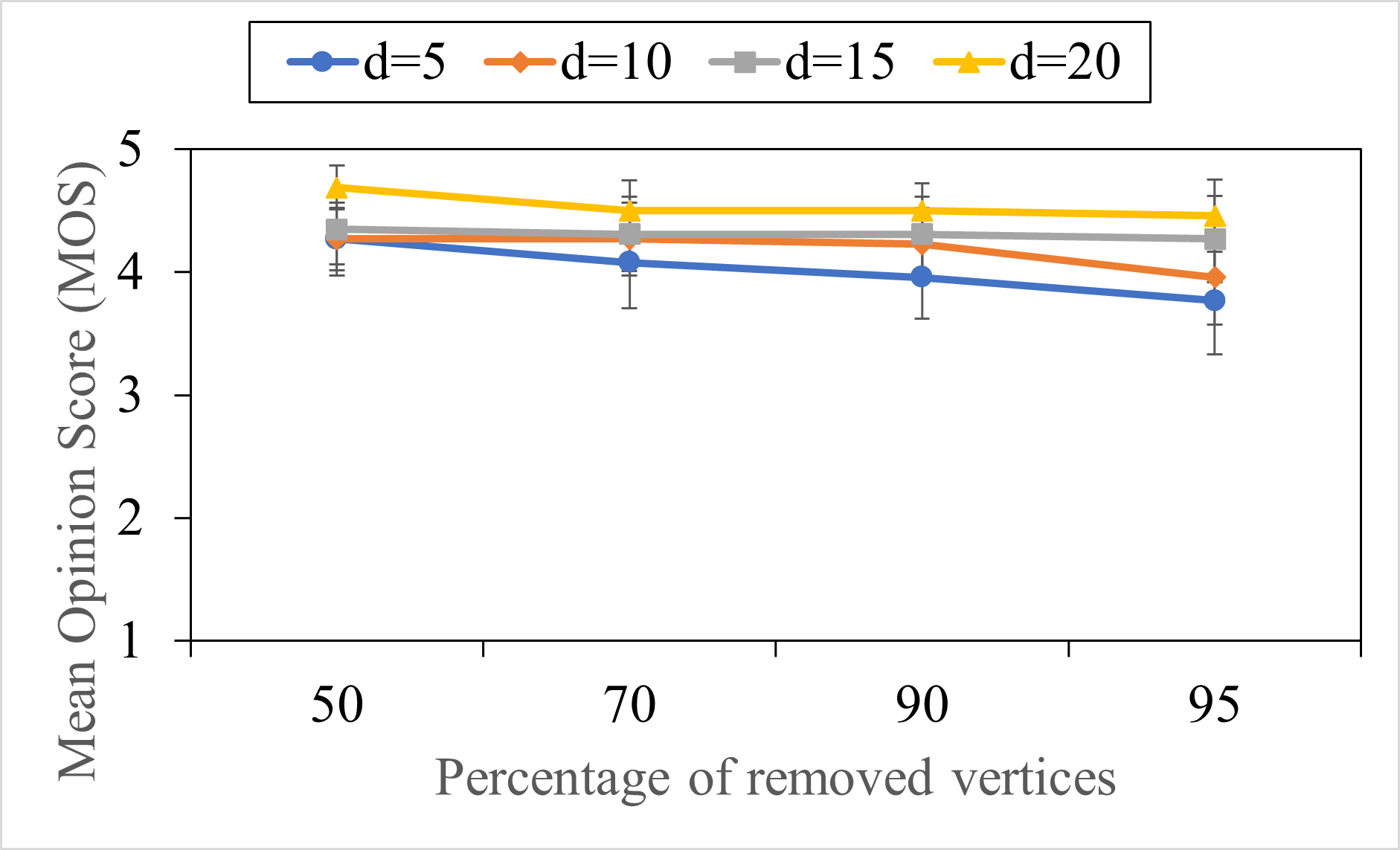}
    }
    \subfloat[Dwarf]{
        \centering
        \includegraphics[width=0.33\textwidth]{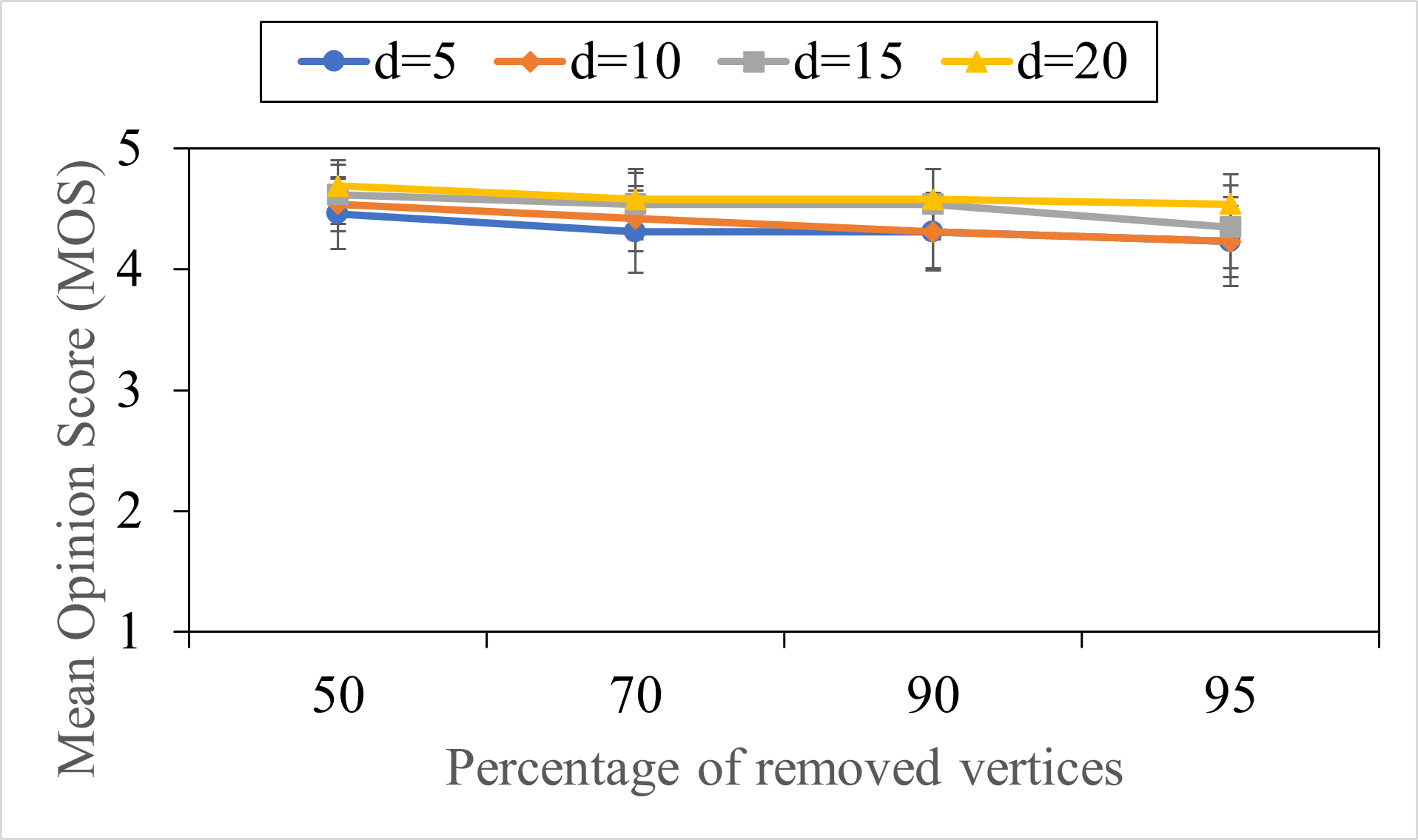}
    }
    \caption{Mean user rating at different LoD versions of the five 3D meshes.}
    \label{fig:LoD_impact}
\end{figure*}

\subsection{Impact of LoD version}
In this part, we analyze the impact of LoD version on the user ratings. Fig.~\ref{fig:LoD_impact} shows the mean user rating of different LoD versions of the five 3D meshes. Here, a LoD version is represented by the percentage of removed vertices. It can be seen that the impact of geometry downsampling varies across the considered meshes. For the Squirrel and the Hulk meshes, the MOS score decreases significantly as the percentage of removed vertices increases across all distance settings. The user rating of the \textit{Squirrel} and \textit{Hulk} models drops by $2.0\sim3.0$ MOS and $2.0\sim3.0$ MOS, respectively. It can also be noted that the downtrend of the user rating is quite consistent at all distance settings. For the \textit{Statue} mesh, the reduction in MOS score is significant only at the smallest distance of $d=5$. At higher distance settings, the difference between the MOS scores of Version 1 (removed 50\%) and Version 4 (removed 95\%) is less than 0.35. In addition, the MOS scores of Version 4 at distances farther than 5 are higher than 4 MOS. In the case of the \textit{Car} and \textit{Dwarf} meshes, it can be seen that the MOS score decreases very slightly as the percentage of the removed vertices increases. Especially, the difference in MOS core between Version 1 and Version 4 of the \textit{Dwarf} model is less than 0.3.
\begin{figure}[t]
    \subfloat[Number of vertices and average MOS decrease of the fived considered meshes]{
    \centering
    \includegraphics[width=\columnwidth]{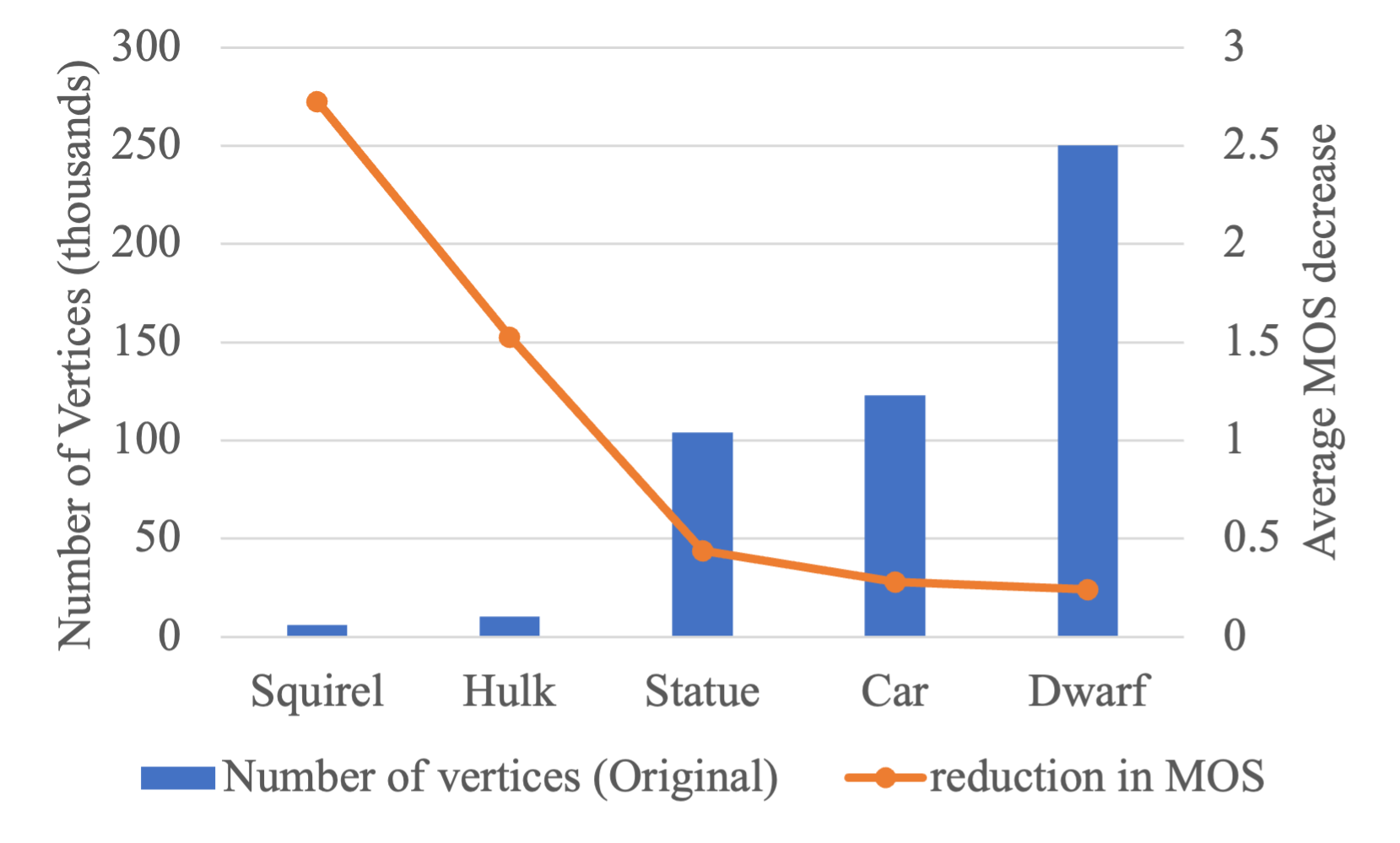}
    }
    \hfil
    \subfloat[Number of vertices and average MOS scores of the fived considered meshes]{
    \includegraphics[width=\columnwidth]{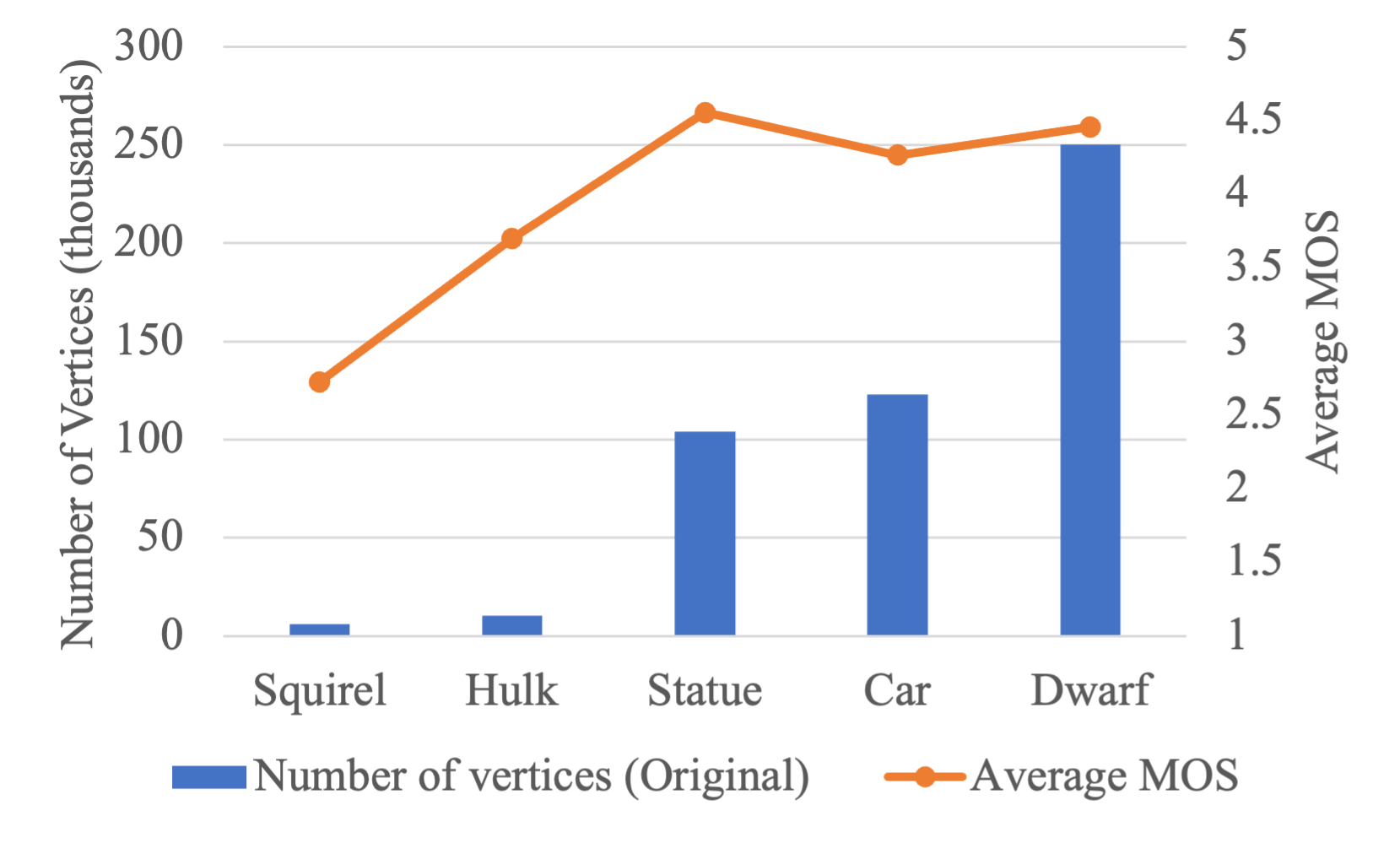}
    }
    \caption{Impacts of content characteristics.}
    \label{fig:content-impact}
\end{figure}
\subsection{Impact of Content Charatertistics}
In this part, we analyze the impact of mesh characteristics on the user rating.
Figure~\ref{fig:content-impact}a plots the average decrease in MOS (Version 4 vs. Version 1) against the number of vertices of the original meshes. It can be seen that the impact of geometry downsampling on the MOS score has a strong correlation with the number of vertices of the meshes. In particular, the smaller the number of vertices of the original model is, the bigger the reduction level in the MOS score becomes. For those meshes with the number of vertices higher than 100K, the difference in the MOS scores between the LoD versions is relatively small. Figure~\ref{fig:content-impact}b shows the relationship between the average MOS score and the number of vertices of the original mesh. It can be seen that the mesh with a higher number of vertices tends to have a higher average MOS. Especially, the average MOS score of the \textit{Statue, Car, and Dwarf} meshes is higher than four.
\begin{table}[t]
\caption{Model parameters and coefficient of determination ($R^2$) with respect to three meshes.}
\label{tab:vertex_num1}
\begin{tabular}{cccccl}
 \toprule
\textbf{Mesh} & 
\textbf{$\alpha$} & 
\textbf{$\beta$} & \textbf{$\gamma$} & \textbf{$\delta$} & \textbf{$R^2$} \\ 
\midrule
Squirrel & 1.2818 & 2.2571 & -0.2534 &-6.3880 & 0.99 \\ 
Hulk & 1.0466 & -0.3993 & 0.1123 &-6.26746 & 0.99 \\ 
Statue & 0.7676 & 2.8189 & -0.2415 & -4.0358 & 0.97 \\ 
Car & 0.3862 & 1.4810 & -0.1146 & -0.3960 & 0.89 \\
Dwarf & 0.1249 & 0.3506 & -0.0149 & 2.6648 & 0.84\\
 \bottomrule
 \end{tabular}
\end{table}

\begin{table}[t]

\caption{Performance of the proposed model in terms of  Pearson correlation coefficient(PLCC), Spearman's rank correlation coefficient (SROCC), and Root mean squared error(RMSE).}
\label{tab:model_performance}
\begin{tabular}{cccl}
 \toprule
Mesh  & PLCC & SROCC & RMSE \\ 
 \midrule
Squirrel & 0.99 & 0.99 & 0.08 \\ 
Hulk & 0.99 & 1.00 & 0.10 \\ 
Statue  & 0.98  & 0.98  & 0.06 \\ 
Car & 0.94 & 0.98 & 0.07 \\
Dwarf & 0.92 & 0.92 & 0.06 \\
 \bottomrule
 \end{tabular}
\end{table}

\begin{figure}[t]
    \includegraphics[width=\columnwidth]{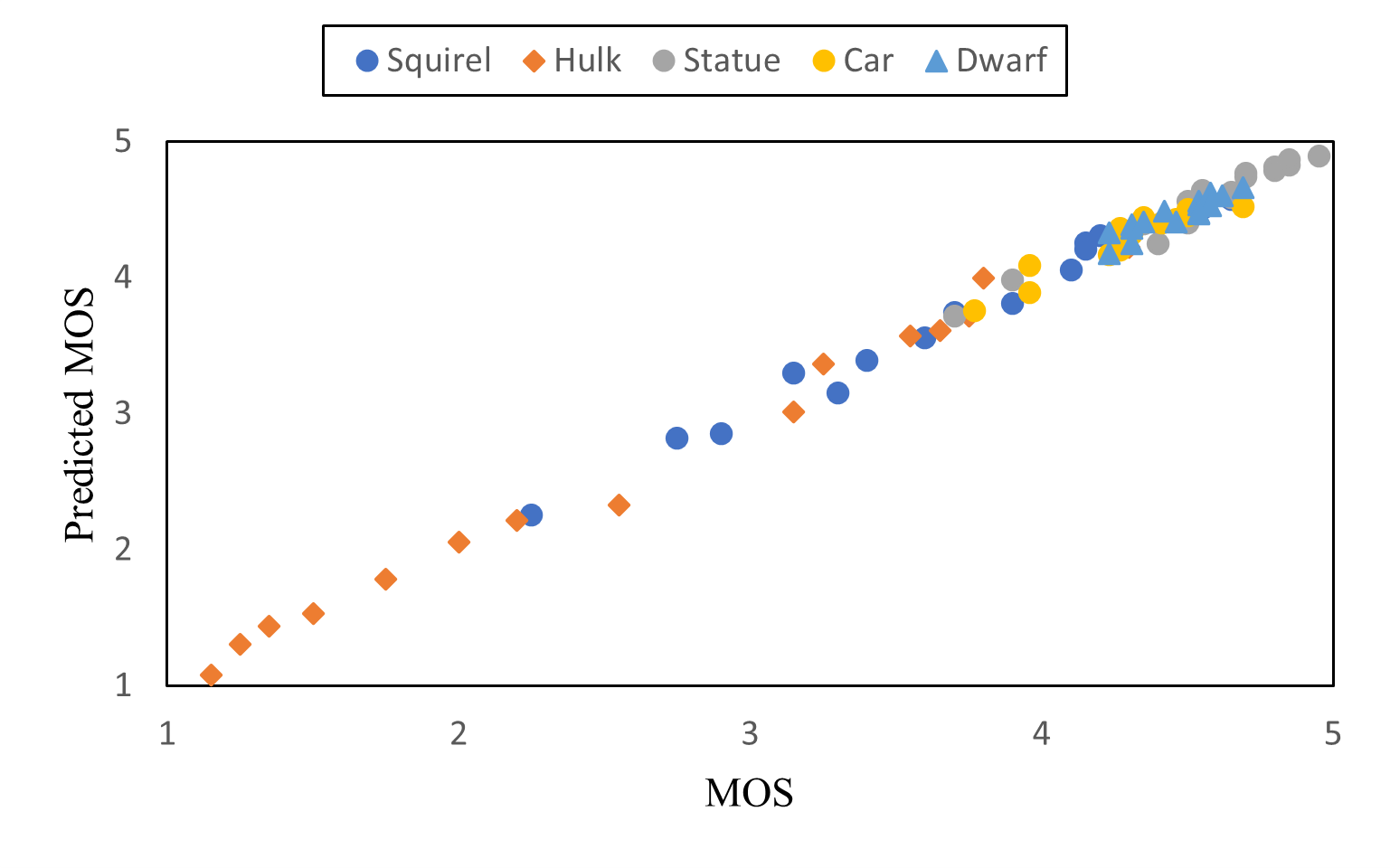}
    \caption{Scatter diagram of the predicted MOS and the subjective scores.}
     \label{fig:scatter}     
\end{figure}

\section{3D Mesh Quality Model}
In this part, we model the subjective scores as a function of the distance from the viewer and the number of vertices of the LoD versions. In particular, we propose to model the MOS scores with the following function.
\begin{equation}
    \textit{MOS} = \alpha\times ln(Q)+ \beta\times ln(D) + \gamma\times ln(Q) \times ln(D) + \delta
\end{equation}
Here, $Q$ denotes the value of the vertices of a LoD version to that of the original model. $D$ denotes the distance to the viewer. $ln(\cdot)$ denotes the natural logarithm function. In the proposed model, the first term represents the impact of the number of vertices on perceptual quality. The second term represents the influence of the distance to the viewer. The mutual impact of the distance and the number of vertices is represented by the third term.  $\alpha, \beta, \gamma$, and $\delta$ are model parameters. The model parameters are determined by means of linear regression. The values of the parameters and coefficient of determination of the proposed model are shown in Table~\ref{tab:vertex_num1}. The scatter diagram of the predicted MOS values and the subjective scores is shown in Fig.~\ref{fig:scatter}. It can be seen that the proposed model achieves high values of the coefficient of determination for all three meshes ($R^2 \geq 0.97$). The performance of the proposed model in terms of Pearson correlation coefficient (PLCC), Spearman's rank correlation coefficient (SROCC), and Root mean squared error(RMSE) is shown in Table~\ref{tab:model_performance}. We can see that the PLCC and SROCC values are higher than 0.98 for all three meshes, while the RMSE values are smaller than 0.10. This result indicates that the proposed model can accurately predict the subjective scores of the LoD versions.

\section{Conclusions}
In this paper, we have studied the user perception of 3D meshes with dynamic Level of Detail in Virtual Reality. Our study shows that the user perception of the mesh quality is affected by not only the level of detail but also the position of the mesh in the virtual environment. In addition, the impact of geometry downsampling is largely dependent on the number of vertices of the original mesh. For the meshes with 100K+ vertices, it is possible to remove up to 95\% number of vertices without significant impact on the user-perceived quality. In future work, we will study the method to generate optimal sets of LoD versions for a given mesh.

\bibliographystyle{IEEEbib}
\bibliography{strings,refs}

\end{document}